# High-precision spectral tuning of micro and nanophotonic cavities by resonantly enhanced photoelectrochemical etching

Eduardo Gil-Santos,[1] Christopher Baker,[1] Aristide Lemaître,[2] Carmen Gomez,[2] Sara Ducci,[1] Giuseppe Leo,[1] and Ivan Favero [1,*]

[1] Matériaux et Phénomènes Quantiques, Université Paris Diderot, CNRS UMR 7162, Sorbonne Paris-Cité, 10 rue Alice Domon et Léonie Duquet, 75013 Paris, France
[2] Laboratoire de Photonique et de Nanostructures, Route de Nozay, CNRS, 91460 Marcoussis, France

*Corresponding author: ivan.favero@univ-paris-diderot.fr

**We present a simple method to tune optical micro- and nanocavities with picometer precision in the resonant wavelength, corresponding to an effective sub atomic monolayer control of the cavity dimension. This is obtained through *resonant* photo-electrochemical etching, with in-situ monitoring of the optical spectrum. We employ this technique to spectrally align an ensemble of resonant cavities in a permanent manner, overcoming the dimension variability resulting from current nanofabrication techniques. In a device containing several resonators, each is individually addressed and tuned, with no optical quality factor degradation. The technique is general and opens the way to multiple applications, such as the straightforward fabrication of networks of identical coupled resonators, or the tuning of chip-based cavities to external references.**

The advent of micro- and nanofabrication techniques has had a profound impact on the control of light-matter interaction, be it in semiconductor, metallic or dielectric materials. It is today possible to tailor the optical properties of emitters by engineering their coupling to solid-state resonant cavities, waveguides or antennas. Collective photonic architectures based on micro- and nanoscale resonators serve in great many contexts like information processing [1,2], sensing [3], metamaterials [4], polaritonics [5], non-linear optics [6], optomechanics [7] and plasmonics [8]. Lattices of resonant cavities are also becoming a test-bed for the physics of strongly correlated optical systems [9], with a recent focus on synchronization [10], phase-transition [11] and topological phenomena [12]. However, in all these frontier applications, there is still a gap between ideas and experimental realizations, due to insufficient deterministic control in the resonant optical wavelength of single and multiple resonators, which precludes many of the above concepts to become reality. Technologically speaking, conventional nanofabrication tools like electronic beam lithography or ion milling and etching typically allow for a nanometer precision in the finished devices dimensions at best. For micro- and nanoscale cavities, such imprecision translates into an appreciable uncertainty in the resonant optical wavelength. As an example, let us consider a whispering gallery mode (WGM) disk resonator of nominal radius R=1µm, fabricated out of a strongly refractive semiconductor like Gallium Arsenide (GaAs), and resonating at a wavelength $\lambda_{WGM}$ [13,14]. Since $\delta\lambda_{WGM}/\lambda_{WGM}=\delta R/R$ [13], with $\delta R$ the radius imprecision, the near-infrared optical modes of such disk, once fabricated, also face an imprecision $\delta\lambda_{WGM}$ of a few nanometers in their resonance. For two resonators to share a common resonance frequency, a precision $\delta R<R/Q$ is required, with Q the resonator quality factor. Even with a modest $Q=10^4$, this means a resonator size control at the level of the Angstrom, the level of a single atom. In consequence, two nominally identical cavities, once fabricated, always resonate at distinct wavelengths, precluding collective spectral alignment or resonant interaction with targeted references. This disorder is a major lock for the future of nanophotonics. It needs to be cracked, ideally with a technique simple enough to spread from basic research experiments to industrial settings, where complex photonic architectures will demand reasonable technological complexity in the handling.

Several techniques were proposed to partially address these issues. Nitrogen gas adsorption allowed finely tuning a microdisk optical resonator [15], yet with non-permanent results and at the price of vacuum and cryogenic operation. The tuning of a photonic crystal cavity was obtained by deposition of a photochromic film [16] or functionalization with a laser-addressable poly-electrolyte [17]. However, these techniques complicate the handling of multiple cavities and tend to degrade optical and mechanical properties by surface modification. Electrical tuning was demonstrated in silicon micro-disks [18] and micro-rings [19], but it implies electrical power consumption and contacts on each resonator, with a presumable impact on integration. The optical tuning of silica microspheres was obtained under exertion of a mechanical stress [20], an approach that lacks scalability and compatibility with on-chip photonics. Gallium Nitride resonators were photo-etched under a ultra-violet spot in a liquid [21], however with a tuning precision inferior to the above methods. For the latter technique to be site-specific, the illumination beam should spatially target each individual resonator. The thermal tuning of silicon [22] and silicon nitride [23] micro-resonators was also demonstrated, but unfortunately comes with continuous energy consumption, non-permanent effects and poor scalability.

Here we present a simple method to achieve spectral tuning of micro- and nanophotonic resonant cavities, which solves all of the above shortcomings. The method uses the concept of *resonant* photo-electrochemical (PEC) etching, whereby a dielectric material forming the optical cavity is immersed in a liquid and etched only in the presence of light resonating in the cavity. The technique requires little equipment, produces permanent results with no degradation, and is automatically scalable to multiple resonators. It is both site-specific, in that it resonantly addresses each individual resonator of an ensemble, and collective, because it does not require individual identification of each resonator to achieve spectral alignment of the ensemble. Being adjustable by light intensity, it allows both fine-tuning of the resonator's wavelength at a picometer level, and coarse-tuning over tens of nanometers. To establish the performance of the technique, we first investigate the tuning of an individual resonator, and then demonstrate the collective tuning of an ensemble of three semiconductor photonic cavities immersed in water.

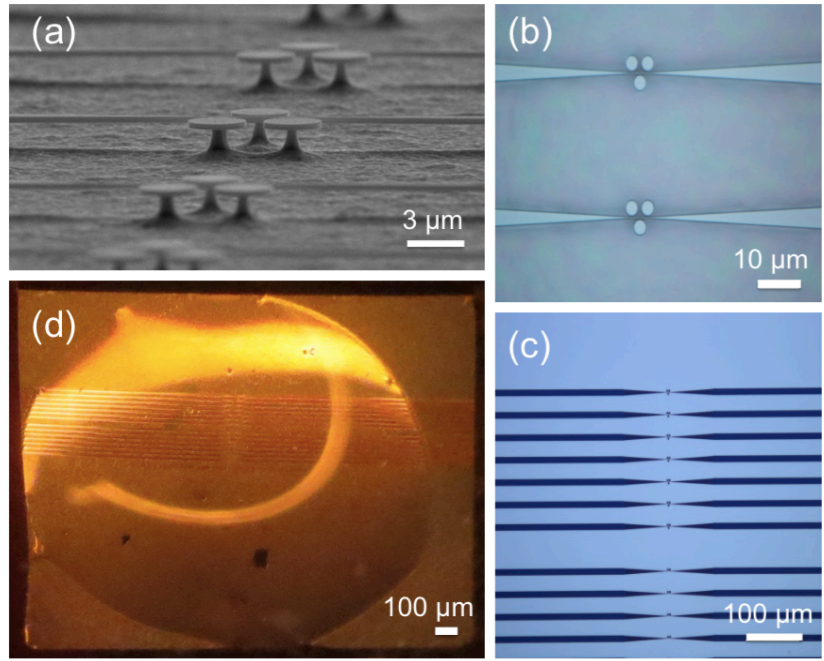

Fig. 1. On-chip WGM GaAs disk resonators. (a) Side-view electron micrograph of sets of 3 disks positioned on both sides of a common GaAs suspended optical-coupling waveguide. (b) Top-view optical micrograph of similar sets, with the tapering of the waveguide made visible. (c) Top-view optical micrograph of the waveguides. (d) Top-view optical micrograph of the chip immersed in a liquid droplet.

The results reported in the following are obtained with GaAs disk resonators, whose typical dimensions are of 320 nm in thickness and 1 μm in radius [24,25]. These resonators support WGMs of mode volume smaller than 1μm$^3$, and represent archetypes of micro/nano photonic cavities for the purposes discussed here. Since photo-electrochemical processes occur in many dielectric semiconductors, the method introduced below in the GaAs environment is in principle transposable to numerous other material platforms, including the important case of silicon. It naturally extends as well to different cavity architectures such those based on photonic crystals and plasmonic designs.

Our disk resonators are isolated from the substrate through an Aluminum Gallium Arsenide (AlGaAs) pedestal, as shown in Fig. 1(a). They are fabricated out of a GaAs (320 nm)/Al$_{0.8}$Ga$_{0.2}$As (1800 nm)/GaAs epitaxial wafer, by electron beam lithography, non-selective dry etching in an Inductively Coupled Plasma Reactive Ion Etching (ICP-RIE) system, and selective under-etching in a dilute hydrofluoric acid (HF) solution. GaAs waveguides integrating a taper with nanoscale transverse dimensions allow evanescent optical coupling of light into the disk's WGMs [26], and are suspended in the resonator's vicinity (see Fig. 1(a) and (b)). A typical chip is millimeter-sized and contains numerous waveguides/disk units (see Fig. 1(c)). Thanks to the large refractive index of GaAs, the chip can be immersed in a transparent liquid while leaving the optical properties of the resonators and waveguides essentially unaffected. The presence of a microliter droplet of water deposited on the sample's surface with a micropipette (see Fig. 1(d)), covering many resonators during tens of minutes before evaporation, still allows optical evanescent coupling experiments to be performed in-situ [27,28]. To that purpose, the continuous-wave monochromatic near-infrared light of a tunable laser is injected into the waveguide at the chip facet [26], and collected at its output on a photodetector (see Supplements). The disk WGMs appear as resonant dips in the waveguide's transmission as the laser wavelength is swept across the spectrum (see Fig. 2). In the upper panel of Fig. 2, the line width of both visible resonances is 32 pm in ambient conditions before liquid immersion, corresponding to a loaded WGM quality factor Q of 41 000. At resonance, the optical power circulating in the cavity is enhanced with respect to the power travelling in the waveguide. As a consequence, once immersed in water, the photo-electrochemical etching of the cavity gets resonantly favored over the etching of the waveguide. The etching reduces the size of the disk resonator and blue shifts its optical resonances. This shift can be continuously controlled and monitored in time in the liquid, by continuously sweeping the laser wavelength over the resonance. A real-time video of such continuous monitoring is shown in the Supplements. Once the desired amount of etching is reached, the laser is switched-off and the sample dried. The dry resonator optical spectrum, whose overall structure is preserved, reveals a permanent shift of 4 nm (see Fig. 2 lower panel). The line width of the two resonances is now of 19 and 25 pm respectively, with an almost unaltered contrast. The loaded Q rises to 70000 and 52000, which represents a clear reduction of optical losses with respect to the situation before etching. We carried out similar tests on many resonators, showing that the optical Q is sometimes improved by the resonant PEC etching, and is at least never degraded. Let us now discuss the mechanisms at play in our tuning technique that are responsible for these advantageous features.

Commonly, photo-electrochemical (PEC) etching consists in optically pumping, above the bandgap, a semiconductor immersed in an electrically conductive liquid. One typically uses ultra-violet light in order to generate free carriers that lead to the formation of ionic species of the semiconductor. The latter are dissolved in the presence of ions provided by the electrolyte liquid [29-31]. In our application, the corresponding chemical reaction for GaAs is $GaAs \rightarrow Ga^{3+} + As^{3+} + 6e^{-}$, whose products react with $OH^{-}$ hydroxyde ions to generate oxides that are then removed. Remarkably, here we achieve PEC etching with light of frequency below the bandgap, in the transparency region of the material.

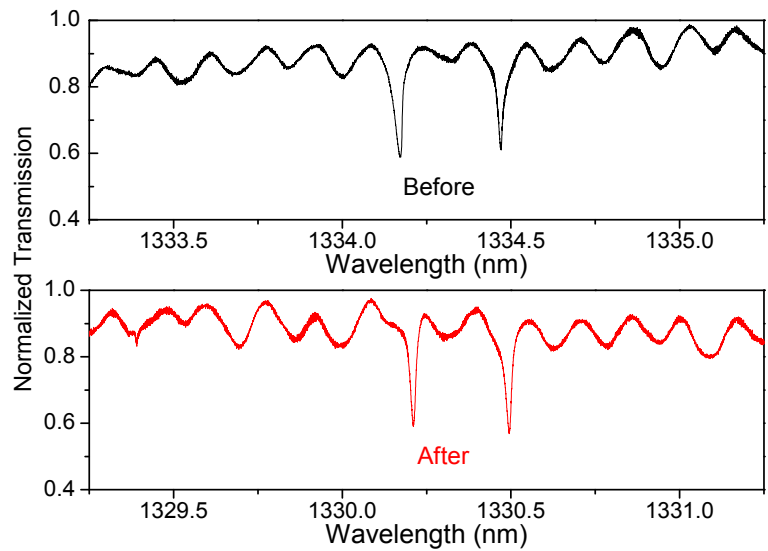

Fig. 2. Resonant PEC shift of the optical spectrum of a GaAs WGM disk resonator. Upper panel: Optical spectrum before resonant PEC etching, with the doublet associated to the two normal modes consisting of a mixture of clockwise and counter-clockwise WGMs [14]. Lower panel: Optical Spectrum after 4 nm of resonant PEC etching. The low-contrast oscillations are residual interferences produced by reflections at the waveguide input and output facets. They can be mitigated in dedicated devices.

Indeed high-Q GaAs disk resonators, and equivalent semiconductor photonic cavities, always absorb a fraction of the resonating light because of residual linear absorption, which in GaAs disks is notably caused by mid-gap states at the resonator's surface [28]. Therefore, the free carriers required for PEC can be generated while pumping below the bandgap. This enables the PEC etching process to be enhanced by the high-Q cavity resonance, which is a central novelty of the here-reported method. This *resonant* nature of our PEC technique enables high spectral selectivity, allowing both to address each individual resonator in a spectrally inhomogeneous ensemble, and high spatial selectivity, by triggering the etching precisely within the optical mode volume of the resonator. The etching process spatially derives from the optical mode profile. In the present case of a WGM, this may smooth out geometric irregularities of the disk resonator, which we anticipate is responsible for the Q improvement sometimes observed after resonant PEC tuning. A second advantage induced by spatial selectivity is that a small flux of light can be used to target each resonator, allowing a high level of control in the etching.

This is exemplified in Fig. 3, which demonstrates the spectral accuracy of the resonant PEC technique by employing a series of elementary fine-tuning cycles. Each cycle consists in rapidly sweeping the wavelength of the laser back and forth over the WGM resonance, which allows both acquiring an optical spectrum and step-tuning the resonating wavelength. The upper panel of Fig. 3 shows the resonance wavelength measured on a GaAs disk WGM as a function of the number of applied cycles. The evolution is linear in the number of cycles, with a blue shift of 7.2 pm per cycle. For the disk considered above, and translated into an effective change of radius, this represents a size control with a precision of 8pm per cycle, which is well below the material's interatomic distance. More precisely, each of these sweep cycles reduces the disk's size by less than a 30th of an atomic monolayer (280 pm). In contrast, simply immersing the chip in an acidic solution to remove the native surface oxide would shift the WGM optical resonance by approximately 2 nm [28,32]. The technique outlined here is more than 250 times more precise, with the added benefit of being able to individually address specific resonators, as we will discuss below.

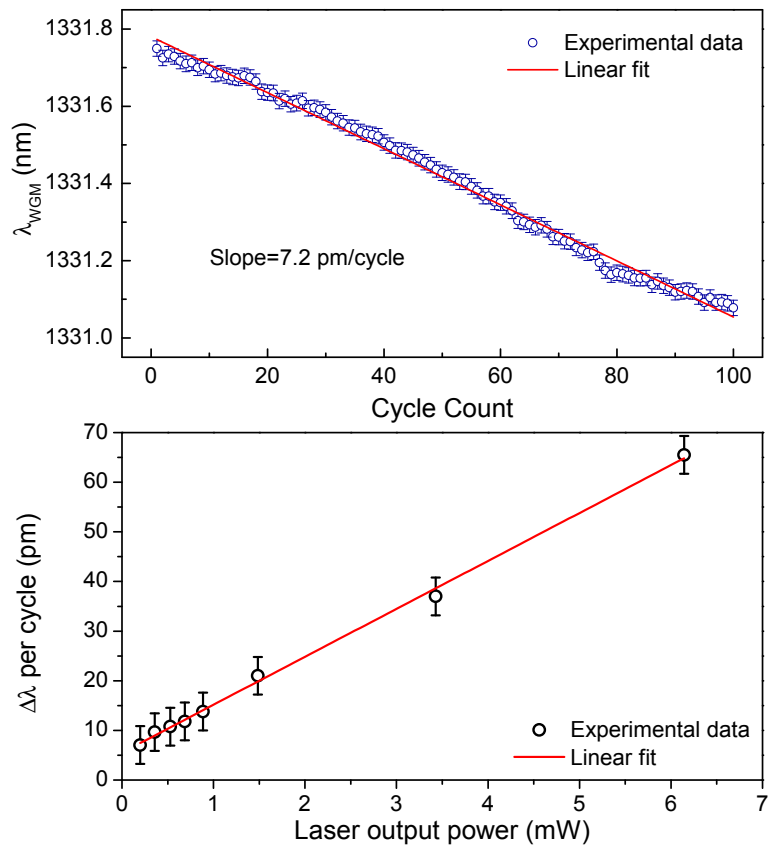

Fig. 3. Fine tuning of a photonic resonator by resonant PEC etching. Upper panel: Wavelength tuning of the WGM of a GaAs disk resonator by subsequent cycles of PEC etching at low optical power, reaching picometer precision. At this level of power, each cycle corresponds to about 1 nJ of energy dropped into the WGM. This would translate into an etch speed of 8 nm/s for a continuous dropped power of 1μW. Lower panel: Wavelength tuning per cycle (see text) as a function of the laser optical output power, which is proportional to the optical power guided in the coupling waveguide. Each point has been obtained by averaging over 100 cycles.

By increasing the laser power, one can adjust the amount of tuning per cycle in a linear fashion, as shown in the lower panel of Fig. 3. By choosing the light intensity and the number of cycles, the resonance wavelength can be tuned between a few picometers and several tens of nanometers, with permanent results. This great spectral versatility of the technique also comes with other benefits when ensembles of resonating cavities are dealt with, as we illustrate thereafter.

The upper panel of Fig. 4 shows a series of nine optical spectra acquired on a set of three distinct disks placed around the same optical coupling waveguide, in the configuration shown in Fig. 1. (a) and (b). The first (lower) spectrum contains three doublets, each doublet being associated to a distinct disk. Even though the three disks are nominally the same, they optically resonate at distinct wavelengths, because of nanofabrication tolerances. In the case shown here, Disk 1 resonates at the largest wavelength, around 1339 nm. By tuning the laser to this wavelength, we selectively trigger the resonant PEC etching in Disk 1 and consequently blue shift its resonance. A second spectrum is then acquired (step 2), revealing this shift, while the two other disks remain barely affected. This procedure is then repeated, to bring step by step the resonances of Disk 1 spectrally close to those of Disk 2. At step number 8 (see Fig. 4 lower panel), the resonances of Disks 1 and 2 start merging. In the following step, the laser etches these two disks concomitantly, and drags spectrally their common resonances towards those of the third disk (Disk 3).

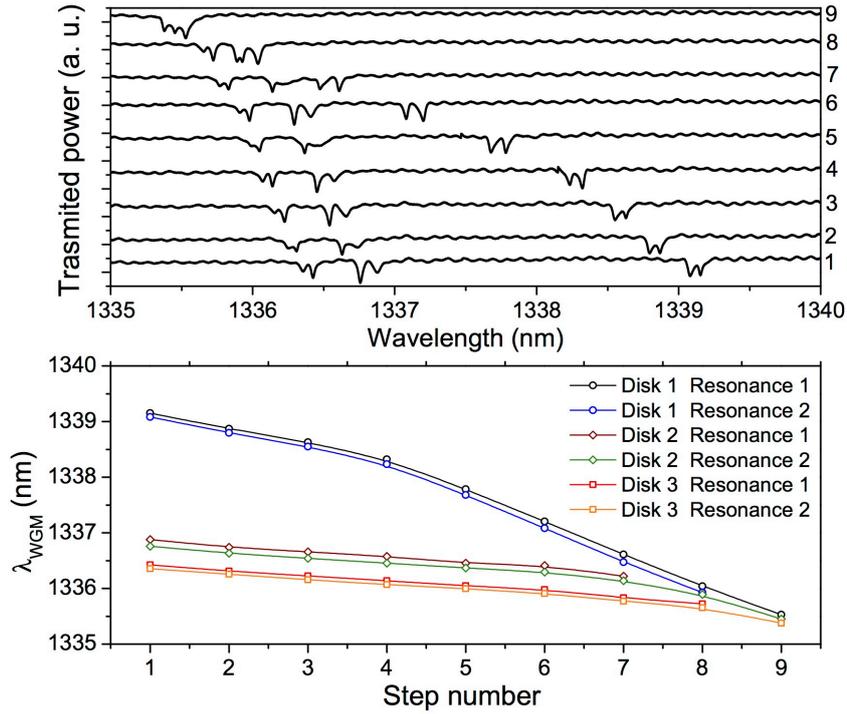

Fig. 4. Collective tuning of three photonic resonators by resonant PEC etching. Upper panel: Optical spectra corresponding to step-by-step spectral alignment of three WGM resonators (see text for details). Lower panel: Follow-up of the WGM resonance wavelengths as a function of the tuning step, showing spectral merging.

At step 9, the optical resonances of the three disks are superimposed in a permanent manner, and the laser can be switched off. The final result is a set of three spectrally aligned resonators. The complete PEC collective tuning procedure is summarized in the Supplements (Fig. S3). Important is to note that, while nine intermediate scans are shown in Fig. 4 in order to illustrate the process, the PEC spectral alignment can also be performed by a single slow continuous laser sweep starting from the red detuned side of resonances. This continuous collective tuning mode is illustrated by a video in the Supplements.

The above collective tuning method can be scaled to more than three resonators, because spectral alignment does not require singling out each resonator. The number of resonators that can be aligned, just as the accuracy of the alignment, are not limited by the operating principles of the method, and can be improved beyond what is reported here. As detailed in the Supplements, the tuning precision of a few picometers reached above is for example set by the absolute spectral inaccuracy of our laser, and could hence be improved. Since we observed resonant PEC etching of GaAs disks in ammonia and isopropanol as well, we anticipate that other ionic liquids or high humidity atmospheres could be employed to gain even better control on the technique. Specific optimization will be needed for each semiconductor showing PEC capability, like silicon, and for metals and insulators. But overall, because the resonant PEC etching method is site-specific, precise, naturally collective, and gives permanent effects, it may become a powerful tool for nanoscale photonic devices, and motivate extensions to other materials and structures. The field of nanophotonics will continue demanding technological advances of the kind to hold its great promises of new science and applications.

**REFERENCES**

1. J. K. Poon, L. Zhu, G. A. DeRose, and A. Yariv, Opt. Lett. **31**, 456 (2006).
2. M. T. Hill, H. J. S. Dorren, T. de Vries, X. J. M. Leijtens, J. H. den Besten, B. Smalbrugge, Y. S. Oei, H. Binsma, G. D. Khoe, and M. K. Smit, Nature **432**, 206 (2004).
3. A. L. Washburn, M. S. Luchansky, A. L. Bowman and R. C. Bailey, Anal. Chem. **82** (1), 69 (2010).
4. V. M. Shalaev, Nature Photonics **1**, 41 - 48 (2007).
5. T. Jacqmin, I. Carusotto, I. Sagnes, M. Abbarchi, D. Solnyshkov, G. Malpuech, E. Galopin, A. Lemaître, J. Bloch, A. Amo, Phys. Rev. Lett. **112**, 116402 (2014).
6. A. Le Boité, G. Orso, and C. Ciuti, Phys. Rev. Lett. **110**, 233601 (2013).
7. G. Heinrich, M. Ludwig, J. Qian, B. Kubala, and F. Marquardt. Phys. Rev. Lett. **107**, 043603 (2011).
8. Z. Liu, A. Boltasseva, R. H. Pedersen, R. Bakker, A. V. Kildishev, V. P. Drachev, V. M. Shalaev, Metamaterials **2**, 45 (2008).
9. M. Zhang, G. S. Wiederhecker, S. Manipatruni, A. Barnard, P. McEuen, and M. Lipson, Phys. Rev. Lett. **109**, 233906 (2012).
10. C. Liu, A. Di Falco, and A. Fratalocchi, Phys. Rev. X **4**, 021048 (2014).
11. V. Peano, C. Brendel, M. Schmidt, and F. Marquardt, Phys. Rev. X **5**, 031011 (2015).
12. L. Ding, C. Baker, P. Senellart, A. Lemaitre, S. Ducci, G. Leo, and I. Favero, Phys. Rev. Lett, **105**, 26, 263903 (2010).
13. J. C. L. Ding, C. Baker, A. Andronico, D. Parrain, P. Senellart, A. Lemaître, S. Ducci, G. Leo, and I. Favero, "Gallium arsenide disk optomechanical resonators," in *Handbook of Optical Microcavities* (PanStanford, 2014).
14. K. Srinivasan and O. Painter, Appl. Phys. Lett. **90**, 3, 031114 (2007).
15. D. Sridharan, E. Waks, G. Solomon, and J. T. Fourkas, Appl. Phys. Lett. **96**, 15, 153303 (2010).
16. K. Piegdon, M. Lexow, G. Grundmeier, H.-S. Kitzerow, K. Pärschke, D. Mergel, D. Reuter, A. Wieck, and C. Meier, Opt. Exp. **20**, 6, 6060 (2012).
17. J. M. Shainline, G. Fernandes, Z. Liu, and J. Xu, Opt. Exp. **18**, 14, 14345 (2010).
18. Y. Shen, I. B. Divliansky, D. N. Basov, and S. Mookherjea, Opt. Lett. **36**, 14, 2668 (2011).
19. W. von Klitzing, R. Long, V. S. Ilchenko, J. Hare, and V. Lefèvre-Seguin, Opt. Lett. **26**, 3, 166 (2001).
20. N. Niu, T.-L. Liu, I. Aharonovich, K. J. Russell, A. Woolf, T. C. Sadler, H. A. El-Ella, M. J. Kappers, R. A. Oliver, and E. L. Hu, Appl. Phys. Lett. **101**, 16, 161105 (2012).
21. P. Dong, W. Qian, H. Liang, R. Shafiiha, D. Feng, G. Li, J. E. Cunningham, A. V. Krishnamoorthy, and M. Asghari, Opt. Exp. **18**, 19, 20298 (2010).
22. M. Zhang, G. S. Wiederhecker, S. Manipatruni, A. Barnard, P. McEuen, and M. Lipson, Phys. Rev. Lett. **109**, 23, 233906 (2012).
23. L. Ding, C. Baker, P. Senellart, A. Lemaitre, S. Ducci, G. Leo and I. Favero. Appl. Phys. Lett. **98**, 113108 (2011).
24. D. T. Nguyen, C. Baker, W. Hease, S. Sejil, P. Senellart, A. Lemaitre, S. Ducci, G. Leo et I. Favero. Appl. Phys. Lett. **103**, 241112 (2013).
25. C. Baker, C. Belacel, A. Andronico, P. Senellart, A. Lemaitre, E. Galopin, S. Ducci, G. Leo, and I. Favero. Appl. Phys. Lett. **99**, 151117 (2011).
26. E. Gil-Santos, C. Baker, D. T. Nguyen, A. Lemaître, C. Gomez, G. Leo, S. Ducci and I. Favero. Nat. Nanotech. **10**, 810 (2015).
27. D. Parrain, C. Baker, G. Wang, A. Lemaitre, P. Senellart, G. Leo, S. Ducci and I. Favero. Opt. Exp. **23**, 15, 19656 (2015).
28. R. W. Hoisty. J. Electrochem. Soc. 108. 790 (1961).
29. D. Greene. Inst. Phys. Conf. Ser. **33a**. 141 (1977).
30. R. Tenne and G. Hodes, Appl. Phys. Lett. **37**, 428 (1980).
31. A. Badolato, K. Hennessy, M. Atatüre, J. Dreiser, E. Hu, P. M. Petroff, and A. Imamoglu. Science **308**, 5725, 1158 (2005).

# SUPPLEMENTS

### Set-up

Fig. S1 shows a binocular top-view of the set-up employed for the resonant PEC tuning of GaAs disk resonators.

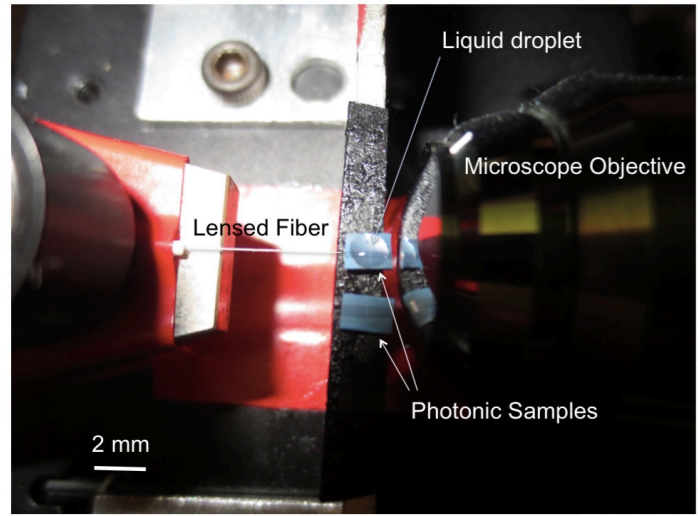

Fig. S1. Resonant PEC etching set-up. Two semiconductor photonic samples are placed on a sample holder (in black). A liquid droplet is formed on the upper sample. This latter sample is sandwiched between an optical micro-lensed fiber for light injection into the sample's waveguides (left) and a microscope objective for light collection at the waveguide's output. The lensed fiber is connected to a tunable external cavity diode laser (Tunics laser) and the light collected by the objective focused back on a p-i-n photodetector. The sample holder, lensed fiber and objective are mounted on xyz micro-positioning stages for alignment.

### Laser imprecision and tuning

The results shown in the main text indicate that a wavelength shift of 7,2 pm per cycle is reached in the resonant PEC tuning process, when subsequent tuning cycles are used at low laser power. This wavelength shift per cycle must be understood as a mean value, as each point of the lower panel of Fig. 3 is obtained by averaging over 100 cycles. Here we show a statistical analysis of the measured per-cycle wavelength shift, to understand the origin of the residual imprecision of our tuning method. The upper panel of Fig. S2 shows a histogram of the shift in a similar low optical power regime, derived from 100 measured cycles performed in water. This histogram shows a mean value of 7 pm for the (negative) wavelength shift. The standard deviation of the measurement amounts to 8 pm, which sets the experimental precision of our cycle tuning mode. The lower panel of Fig. S2 shows the same analysis with the resonators operated in air, hence in absence of PEC etching. The mean value is this time zero (no etching) and the standard deviation 7 pm. This indicates that our cycle-tuning experiments are affected by an imprecision of 7pm in the measurement of the resonant wavelength $\lambda_{WGM}$. We anticipate this imprecision to be associated to a lack of continuous spectral accuracy as we sweep the laser wavelength over subsequent cycles (we use rapid sweep scans of an external cavity diode laser). While this technical aspect seems to currently limit the precision of our resonant PEC tuning in the cycle mode, it is absent in the continuous tuning mode. Simple developments should hence allow improving the (already high) precision shown in the main text.

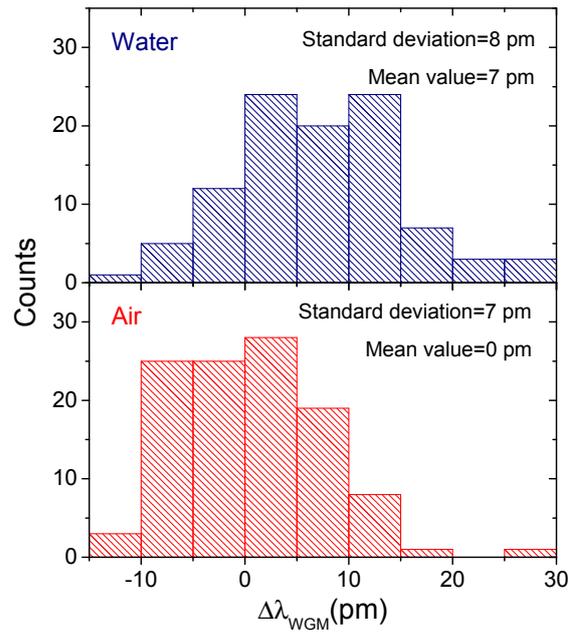

Fig. S2. Laser and tuning precision. Upper panel: Histogram of the measured wavelength shift per cycle in water, built from a series of one hundred cycles of resonant PEC tuning. Lower panel: the same histogram is obtained from measurements performed in air, where no PEC etching occurs.

**Resonant PEC tuning of multiple cavities**

Fig. S3 summarizes the principles of the collective PEC tuning of three detuned resonators.

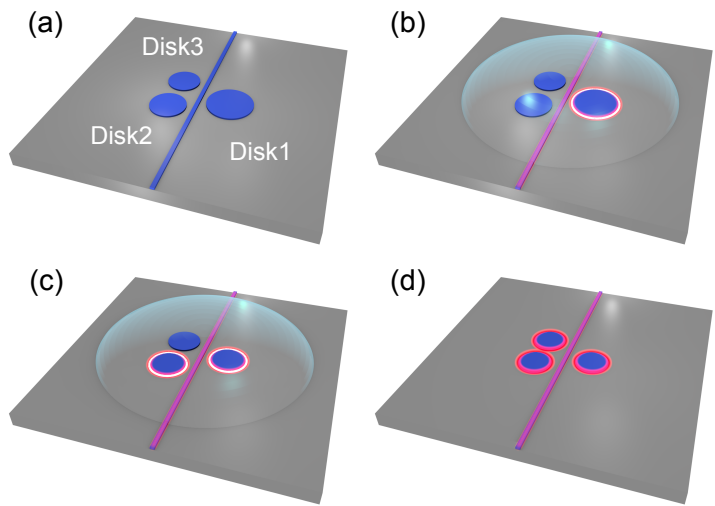

Fig. S3. Principles of the collective PEC tuning. (a) Three disk resonators start with a different size hence different resonance wavelength. (b) The chip is immersed in water. The laser (red color) is turned on and tuned to the WGM of the largest disk, Disk 1, resonating at largest wavelength. Disk 1 is consequently etched (white ring etching zone) and its resonating wavelength progressively blue shifted. (c). After some amount of PEC etching, Disk 1 resonates and the same wavelength as Disk 2, which was the disk of intermediate size, and the two disks get etched together. (d) The final step is reached when Disk 1 and Disk 2 reach the size of Disk 3, which was the smallest disk at the beginning. At this point, the three resonators are spectrally aligned and resonate together.